\documentclass[proceedings, preprint]{rmaa}

\usepackage{paralist}

\usepackage{psfrag,color}

\SetYear{2015}
\SetConfTitle{Fourth Workshop on Robotic Autonomous Observatories}

\title{Meteor observations with Mini-MegaTORTORA wide-field monitoring system}

\author{
  S. Karpov,\altaffilmark{1,3}
  N. Orekhova,\altaffilmark{2}
  G. Beskin,\altaffilmark{1,3}
  A. Biryukov,\altaffilmark{3,4}
  S. Bondar,\altaffilmark{2}
  E. Ivanov,\altaffilmark{2}
  E. Katkova,\altaffilmark{2}
  A. Perkov,\altaffilmark{2}
  and V. Sasyuk\altaffilmark{3}
}

\altaffiltext{1}{Special Astrophysical Observatory of Russian Academy of
  Sciences, Russia.}

\altaffiltext{2}{Research and Production Corporation ``Precision Systems and
  Instruments'', Russia.}

\altaffiltext{3}{Kazan Federal University, Russia.}

\altaffiltext{4}{Sternberg Astronomical Institute, Moscow State University, Russia.}

\shortauthor{Karpov et al.}
\shorttitle{Meteors observations with MMT-9}

\listofauthors{S. Karpov, G. Beskin, A. Biryukov, S. Bondar, E. Ivanov,
  E. Katkova, N. Orekhova, A. Perkov, V. Sasyuk}
\indexauthor{Karpov, S.}
\indexauthor{Orekhova, N.}
\indexauthor{Beskin, G.}
\indexauthor{Biryukov, A.}
\indexauthor{Bondar, S.}
\indexauthor{Ivanov, E.}
\indexauthor{Katkova, E.}
\indexauthor{Perkov, A.}
\indexauthor{Sasyuk, V.}

\abstract{ Here we report on the results of meteor observations with 9-channel
  Mini-MegaTORTORA (MMT-9) wide-field optical monitoring system with high
  temporal resolution. During first 1.5 years of operation more than 90
  thousands of meteors have been detected, at a rate of 300-350 per night, with
  durations from 0.1 to 2.5 seconds and angular velocities up to 38 degrees per
  second.  The faintest detected meteors has the peak brightness about 10 mag,
  while the majority - from 4 to 8 mag. Some of the meteors have
  been observed in BVR filters simultaneously. Color variations along the trail
  for them are determined.  All parameters of detected meteors are published
  online. The database also includes the information on 10 thousands meteors
  detected by our previous FAVOR camera in 2006-2009 years.}

\addkeyword{meteorites, meteors, meteoroids}
\addkeyword{astronomical databases: miscellaneous}

\begin{document}
\maketitle

\label{sec:intro}

\begin{figure*}[!t]
  \makebox[0pt][l]{\textbf{a}}%
  \hspace*{0.9\columnwidth}\hspace*{\columnsep}%
  \textbf{b}\\[-0.7\baselineskip]
  \parbox[t]{\textwidth}{%
     \vspace{0pt}
     \includegraphics[width=0.90\columnwidth]{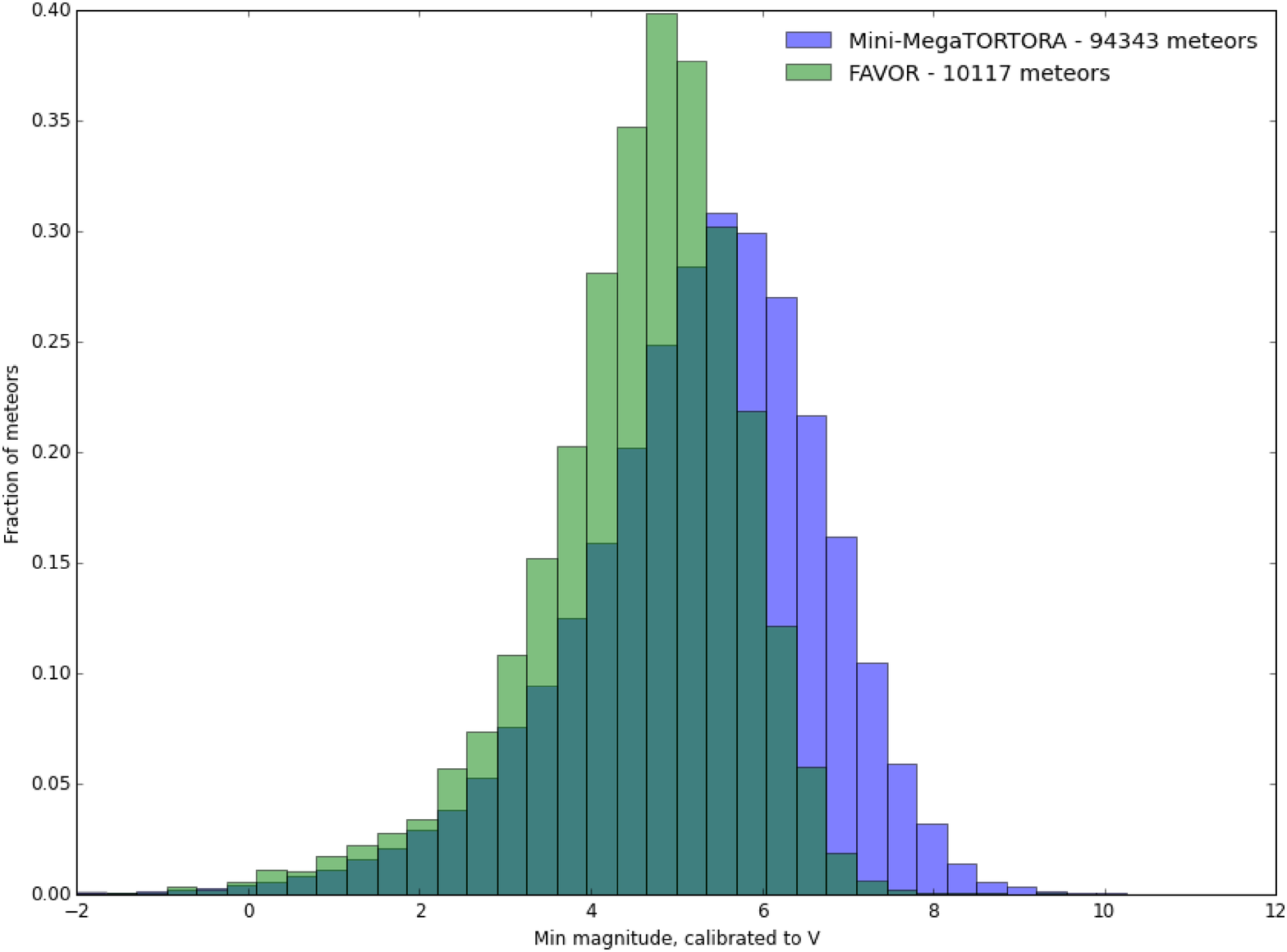}%
     \hfill%
     \includegraphics[width=1.10\columnwidth]{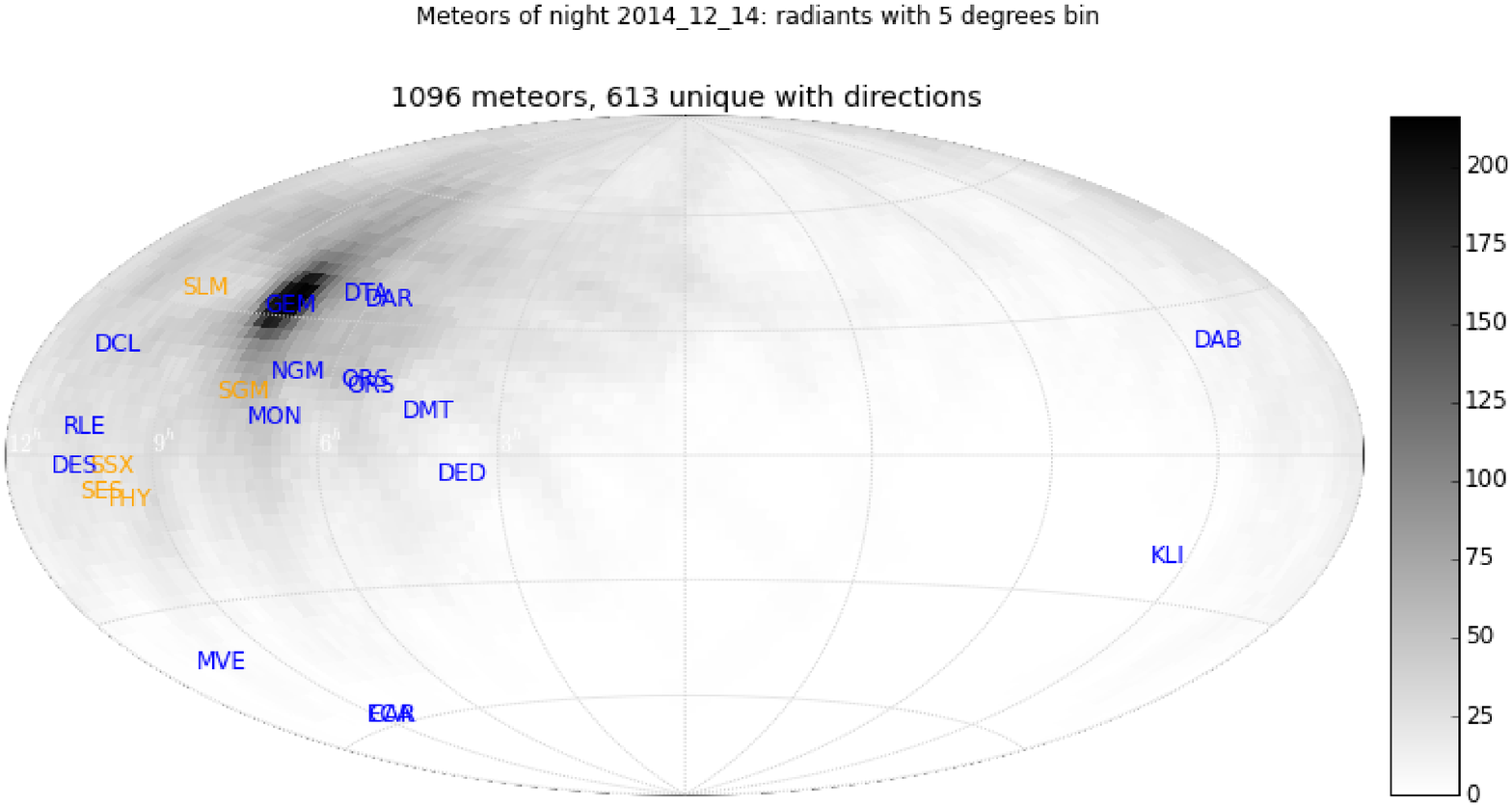}
     }
     \caption{(\textit{a})Peak brightness distribution for meteors detected by
       Mini-MegaTORTORA and FAVOR cameras. (\textit{b}) Density of
       intersections of meteor trails from the night corresponding to the peak
       of 2014 Geminids.
  }
  \label{fig:meteors}
\end{figure*}

Wide-field monitoring systems with sub-second temporal resolution are optimal
instruments to look for and study of rapid transient events of unpredictable
localization, which may be of both cosmological (gamma-ray bursts, supernovae),
Galactic (flaring stars, novae, variable stars) and near-Earth origin (meteors,
asteroids, artificial satellites). FAVOR camera what we developed in early
2000-s \citep{beskin_favor2,favor_malaga} was able to detect a lot of faint
meteors what can't be typically observed using other means. Indeed, typical TV
observations imply a fish-eye cameras with high frame rate, have low
angular resolution and are able to detect only brighter events and fireballs
. On the other hand, ten thousand meteors observed with FAVOR camera from Aug
2006 till Mar 2009, have been significantly, by several magnitudes,
fainter. Unfortunately, these meteors were not published until now and were
mostly unavailable to the analysis by scientific community.


In mid-2014 we started the observations with Mini-MegaTORTORA (MMT-9), which is
a novel multichannel wide-field monitoring camera, placed at Special
Astrophysical Observatory, near Russian 6-m telescope
\citep{beskin_ufn_2010, karpov_acpol_2013, beskin_revmex_2014,
  biryukov_firstlight}.
It continuously monitors the sky with 0.1 s temporal resolution in 900 square
degrees field of view, detecting various kinds of transient events on the fly
using the real-time data processing pipeline. The meteors are extracted by its
elongated shape and all the images containing them, obtained by either one or
several channels, are analyzed automatically to derive its brightness along the
trail, light curve, trajectory, angular velocity and duration. The majority of
events are observed in white light (the brightness is then calibrated to V
magnitude), while some are being observed in Johnson-Cousins B, V and R
photometric filters simultaneously. For such events, the colors are also
derived automatically.  All these data for 94343 (as of Dec 9, 2015) events are
stored to the database and are available
online\footnote{The database is published at
  \url{http://mmt.favor2.info/meteors} and \url{http://astroguard.ru/meteors}}.

The database also contains the same information for 10117 meteors observed with
FAVOR camera in 2006-2009 years, uniformly processed with the same software as
being used for MMT-9 data analysis. Figure~\ref{fig:meteors}a shows the comparison
of peak magnitudes (integral brightness of the meteor trail on a single frame
where the meteor is brightest) of events observed with MMT-9 and FAVOR.  The
faintest detected meteors has the peak brightness about 10 mag, while the
majority -- from 4 to 8 mag, and are much fainter than ones contained in such
meteor databases as SonotaCo\citep{sonotaco} and EDMOND\citep{edmond}.

The database does not presently include any parallactic observation (though we
are working on installing second version of Mini-MegaTORTORA which will allow
us to measure meteor parallaxes). However, huge amount of meteors measured
every night might in principle allow to detect the radiants of meteor streams
using purely statistical methods. Figure~\ref{fig:meteors}b shows the density
of intersections of meteor trails from the night corresponding to 2014
Geminids, and the radiant is clearly visible here.

We hope that the database of meteor observations what we publish will help in
studying faint component of meteor showers.


\section*{Acknowledgements}
This work was supported by the grants of RFBR (No. 09–02–12053 и
12–02–00743-A), by the grant of European Union (FP7 grant agreement number
283783, GLORIA project). Mini-MegaTORTORA belongs to Kazan Federal University
and the work is performed according to the Russian Government Program of
Competitive Growth of Kazan Federal University. Observations on
Mini-MegaTORTORA are supported by the Russian Science Foundation grant
No. 14-50-00043.

\bibliography{meteors}

\begin{thebibliography}
\expandafter\ifx\csname natexlab\endcsname\relax\def\natexlab#1{#1}\fi
\expandafter\ifx\csname href\endcsname\relax
  \def\href#1#2{}\fi
\expandafter\ifx\csname urllinklabel\endcsname\relax
  \def\urllinklabel{[LINK]}\fi
\expandafter\ifx\csname adsurllinklabel\endcsname\relax
  \def\adsurllinklabel{[ADS]}\fi

\bibitem[{{Beskin} {et~al.}(2014){Beskin}, {Karpov}, {Bondar}, {Perkov},
  {Ivanov}, {Katkova}, {Sasyuk}, {Biryukov}, \& {Shearer}}]{beskin_revmex_2014}
{Beskin}, G., {Karpov}, S., {Bondar}, S., {Perkov}, A., {Ivanov}, E.,
  {Katkova}, E., {Sasyuk}, V., {Biryukov}, A., \& {Shearer}, A. 2014, in
  Revista Mexicana de Astronomia y Astrofisica Conference Series, Vol.~45,
  Revista Mexicana de Astronomia y Astrofisica Conference Series, 20--


\bibitem[{{Beskin} {et~al.}(2010){Beskin}, {Karpov}, {Bondar},
  {Plokhotnichenko}, {Guarnieri}, {Bartolini}, {Greco}, \&
  {Piccioni}}]{beskin_ufn_2010}
{Beskin}, G.~M., {Karpov}, S.~V., {Bondar}, S.~F., {Plokhotnichenko}, V.~L.,
  {Guarnieri}, A., {Bartolini}, C., {Greco}, G., \& {Piccioni}, A. 2010,
  Physics Uspekhi, 53, 406


\bibitem[{{Biryukov} {et~al.}(2015){Biryukov}, {Beskin}, {Karpov}, {Bondar},
  {Ivanov}, {Katkova}, {Perkov}, \& {Sasyuk}}]{biryukov_firstlight}
{Biryukov}, A., {Beskin}, G., {Karpov}, S., {Bondar}, S., {Ivanov}, E.,
  {Katkova}, E., {Perkov}, A., \& {Sasyuk}, V. 2015, Baltic Astronomy, 24, 100


\bibitem[{{Karpov} {et~al.}(2005){Karpov}, {Beskin}, {Biryukov}, {Bondar},
  {Hurley}, {Ivanov}, {Katkova}, {Pozanenko}, \& {Zolotukhin}}]{beskin_favor2}
{Karpov}, S., {Beskin}, G., {Biryukov}, A., {Bondar}, S., {Hurley}, K.,
  {Ivanov}, E., {Katkova}, E., {Pozanenko}, A., \& {Zolotukhin}, I. 2005, Nuovo
  Cimento C, 28, 747


\bibitem[{{Karpov} {et~al.}(2010){Karpov}, {Beskin}, {Bondar}, {Guarnieri},
  {Bartolini}, {Greco}, \& {Piccioni}}]{favor_malaga}
{Karpov}, S., {Beskin}, G., {Bondar}, S., {Guarnieri}, A., {Bartolini}, C.,
  {Greco}, G., \& {Piccioni}, A. 2010, Advances in Astronomy, 2010


\bibitem[{{Karpov} {et~al.}(2013){Karpov}, {Beskin}, {Bondar}, {Perkov},
  {Ivanov}, {Guarnieri}, {Bartolini}, {Greco}, {Shearer}, \&
  {Sasyuk}}]{karpov_acpol_2013}
{Karpov}, S., {Beskin}, G., {Bondar}, S., {Perkov}, A., {Ivanov}, E.,
  {Guarnieri}, A., {Bartolini}, C., {Greco}, G., {Shearer}, A., \& {Sasyuk}, V.
  2013, Acta Polytechnica, 53, 38


\bibitem[{{Korno{\v s}} {et~al.}(2014){Korno{\v s}}, {Koukal}, {Piffl}, \&
  {T{\'o}th}}]{edmond}
{Korno{\v s}}, L., {Koukal}, J., {Piffl}, R., \& {T{\'o}th}, J. Proceedings of
  the International Meteor Conference, Poznan, Poland, 22-25 August 2013, ed. ,
  M.~{Gyssens}P.~{Roggemans} \& P.~{Zoladek}, 23--25


\bibitem[{{SonotaCo}(2009)}]{sonotaco}
{SonotaCo}. 2009, WGN, Journal of the International Meteor Organization, 37, 55


\end{thebibliography}

\end{document}